\documentclass[aps, pre, onecolumn, nofootinbib, notitlepage, groupedaddress, amsfonts, amssymb, amsmath, longbibliography]{revtex4-1}
\usepackage{graphicx}
\usepackage{hyperref}
\usepackage{xcolor}
\hypersetup{
    colorlinks,
    linkcolor={red!50!black},
    citecolor={blue!50!black},
    urlcolor={blue!80!black}
}
\usepackage{bm}
\usepackage{natbib}
\usepackage{longtable}
\newcommand{\Div}[1]{\ensuremath{\nabla\cdot\left( #1\right)}}

\newcommand{\grad}{\ensuremath{\nabla}}
\newcommand{\RB}{Rayleigh-B\'{e}nard }

\newcommand{\lilstressT}{\ensuremath{\bm{\bar{\bar{\sigma}}}}}

\newcommand{\approptoinn}[2]{\mathrel{\vcenter{
	\offinterlineskip\halign{\hfil$##$\cr
	#1\propto\cr\noalign{\kern2pt}#1\sim\cr\noalign{\kern-2pt}}}}}

\newcommand{\appropto}{\mathpalette\approptoinn\relax}

\newcommand\mnras{{MNRAS}}%

\begin{document}
\author{Evan H. Anders}
\affiliation{Dept. Astrophysical \& Planetary Sciences, University of Colorado -- Boulder, Boulder, CO 80309, USA}
\affiliation{Laboratory for Atmospheric and Space Physics, Boulder, CO 80303, USA}
\author{Benjamin P. Brown}
\affiliation{Dept. Astrophysical \& Planetary Sciences, University of Colorado -- Boulder, Boulder, CO 80309, USA}
\affiliation{Laboratory for Atmospheric and Space Physics, Boulder, CO 80303, USA}
\title{Convective heat transport in stratified atmospheres at low and high Mach number}

\begin{abstract}
We study fully compressible convection in the context of 
plane-parallel, polytropically stratified atmospheres. 
We perform a suite of 2D and 3D simulations in which we vary the initial
superadiabaticity ($\epsilon$) and the Rayleigh number (Ra) while fixing the
initial density stratification, aspect
ratio, and Prandtl number.
The evolved heat transport, 
quantified by the Nusselt number (Nu),
follows scaling relationships similar to those found in the well-studied, 
incompressible Rayleigh-B\'{e}nard problem.  This scaling holds up in both 2D and 3D and is not
appreciably affected by the magnitude of $\epsilon$.
\end{abstract}
\maketitle

\section{Introduction}
\label{sec:intro}
Convection transports energy in stellar and planetary atmospheres
where flows are compressible and
feel the atmospheric stratification.  This stratification 
is significant in regions such as
the convective envelope of the Sun, which spans 14 density scale heights.
In the bulk of these systems, particularly in the deep interior,
flows are at very low Mach number (Ma).  Unfortunately,
numerical constraints have restricted most studies of 
compressible convection to high Ma.
These prior studies \cite{graham1975, chan&all1982,
hurlburt&all1984, cattaneo&all1990, brummell&all1996,
brandenburg&all2005} have provided insight into the nature of
convection in the low temperature,
high Ma region near the Sun's surface. Few fundamental
properties of low Ma compressible convection, such as the scaling of
convective heat transport, are known.

In the widely-studied \RB problem of incompressible Boussinesq convection (RBC), 
a sufficiently negative temperature gradient causes convective instability.
In the evolved solution, upflows and downflows are symmetrical, the
temperature in the interior becomes isothermal, and
the conductive flux ($\propto \grad T$) approaches 
zero there. 
For compressible convection in a stratified atmosphere, a
negative entropy gradient causes convective instability.
Early numerical experiments of moderate-to-high Ma compressible convection
in two \cite{graham1975, chan&all1982,
hurlburt&all1984, cattaneo&all1990} and three 
\cite{cattaneo&all1991, brandenburg&all2005, brummell&all1996} dimensions
revealed a different evolved state from RBC.
Downflow lanes
become fast and narrow, and upflow lanes turn into broad, slow upwellings.
Furthermore, the \emph{entropy} gradient is negated by convection in the interior, so
a significant temperature gradient and conductive flux can persist despite
efficient convection.

In RBC, there exist two primary dynamical control parameters: 
the Rayleigh number (Ra, the ratio of
buoyant driving to diffusive damping) and the Prandtl number 
(Pr, the ratio of viscous to thermal
diffusivity). These numbers control two useful
measures of turbulence in the evolved solution:
the Reynolds
number (Re, the strength of advection to viscous diffusion)
and the Peclet number (Pe, advection vs. thermal diffusion).  
In stratified atmospheres, the magnitude of the unstable entropy gradient
joins Ra and Pr as a third important and independent control parameter.  This 
\emph{superadiabatic excess}, $\epsilon$,
sets the scale of the atmospheric entropy gradient \cite{graham1975}.
We find here that $\epsilon$ primarily controls the Ma of the evolved solution.

Here we study the behavior of convective heat transport, 
quantified by the Nusselt number (Nu), in plane-parallel, 
two- and three-dimensional, polytropically stratified atmospheres.  
We vary $\epsilon$ and Ra while holding Pr, aspect ratio, boundary conditions,
and initial atmospheric stratification
constant.  We also examine the behavior of flow properties, as quantified by Ma and Re.
We find here that the scaling of Nu in stratified, compressible convection 
is similar to that in \RB convection,
and that this scaling is not appreciably changed by the magnitude of the superadiabaticity.

\section{Experiment} 
\label{sec:experiment}
We examine a monatomic ideal gas with an adiabatic index of
$\gamma = 5/3$ whose equation of state is $P = R\rho T$. This is consistent with the approach used in earlier work 
\cite{graham1975, chan&all1982, brandenburg&all2005,
hurlburt&all1984, cattaneo&all1990, cattaneo&all1991, brummell&all1996} 
and is the simplest stratified extension of RBC.
The atmospheres studied here are initially polytropically stratified,
\begin{equation}
\begin{split}
\rho_0(z) &= \rho_{t}(1 + L_z - z)^m, \\
T_0(z)    &= T_{t}(1 + L_z - z),
\label{eqn:polytrope}
\end{split}
\end{equation}
where $m$ is the polytropic index and $L_z$ is the depth of the atmosphere.
The polytropic
index is set by the superadiabatic excess, $\epsilon = m_{ad} - m$, where
$m_{ad} = (\gamma - 1)^{-1}$ is the adiabatic value of $m$.
The height coordinate, $z$, increases upwards in the range $[0, L_z]$.
Subscript 0 indicates initial conditions and subscript $t$ indicates values
at $z = L_z$.   Stratified atmospheres have a fourth non-dimensional parameter,
the number of density scale heights, $n_{\rho} = \ln\left[\rho_0(z=0)/\rho_t\right]$.  We
specify the depth of the atmosphere, $L_z = e^{n_{\rho}/m} - 1$, by choosing
the initial value of $n_{\rho}$.
Throughout this work we set $n_{\rho} = 3$.    Satisfying hydrostatic
equilibrium sets the value of gravity, $g = RT_t (m + 1)$, which is
constant with depth.  We study atmospheres with aspect
ratios of 4 where both the $x$ and $y$ coordinates have the range $[0, 4L_z]$.
In our 2D cases, we only consider $x$ and $z$.

These domains are nondimensionalized by setting
$R = T_t = \rho_t = 1$ at $z = L_z$.
By this choice, the non-dimensional
length scale is the inverse temperature gradient scale and the 
timescale is the isothermal sound crossing time, 
$\tau_I$, of this unit length.
Meaningful convective dynamics occur on 
timescales of the atmospheric buoyancy time,
$t_b = \tau_I \sqrt{L_z\,m\,c_P/g\,\epsilon\,n_\rho}$, where
$c_P = R \gamma/(\gamma-1) = 2.5$ is the specific heat at constant pressure.

\begin{figure}[t]
\includegraphics[width=\textwidth]{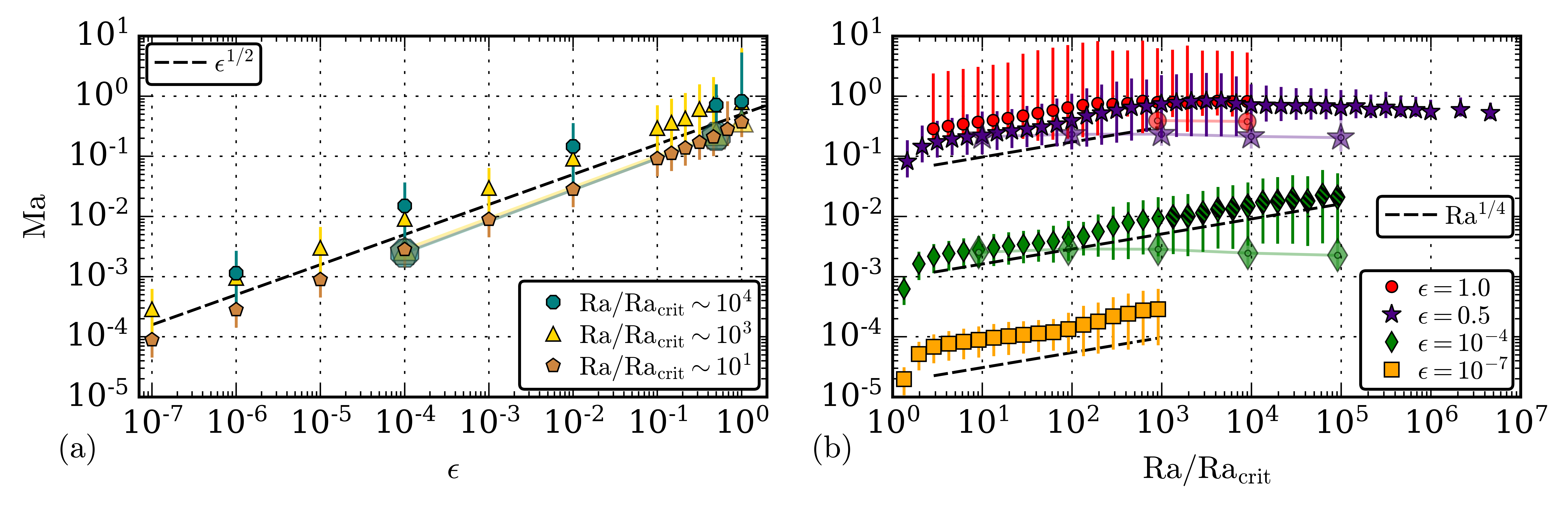}
\caption{The mean adiabatic Mach number of long-time-averaged profiles
is shown.  Error bars show the full range of Ma over the depth of the
atmosphere.
(a) Ma, at various values of Ra/Ra$_{\text{crit}}$, 
is plotted as a function of $\epsilon$.
(b) Ma, at various values of $\epsilon$, 
is plotted as a function of Ra/Ra$_{\text{crit}}$.
Larger symbols with inlaid circles designate 3D runs.
\label{fig:ma_v_eps} }
\end{figure}

At fixed $n_\rho$, convective dynamics are 
controlled by $\epsilon$ as well as the atmospheric diffusivities.
At a fixed value of
$\epsilon$, the diffusivities are set by the
Rayleigh number (Ra) and the Prandtl number (Pr),
\begin{equation}
\text{Ra}_{t} = \frac{g L_z^3 (\Delta S_0 / c_P)}{\nu_t\chi_t},
\qquad
\text{Pr} = \frac{\nu}{\chi},
\end{equation}
where $\Delta S_0 = \epsilon\ln (1 + L_z) = \epsilon n_\rho / m$ 
is the initial specific entropy difference across the domain,
$\chi$ is the thermal diffusivity, 
and $\nu$ is the kinematic viscosity.
Throughout this work we specify
that Pr $= 1$ and is depth invariant.
The initial thermal
conductivity, $\kappa_0 = \chi \rho_0$, is
constant with depth, such that (\ref{eqn:polytrope}) is in
thermal equilibrium ($\grad\cdot[\kappa_0\grad T_0] = 0$).
By these
choices, $\nu(z) \equiv \chi(z) \equiv \chi_t / \rho_0$.
This formulation 
sets Ra at the bottom of the domain greater than
Ra$_t$ by a factor of $e^{2n_\rho}$. Henceforth
when we specify Ra we are referring to Ra$_t$.  
The full values of $\kappa = \rho\chi$ and 
$\mu = \rho\nu$ (the dynamic viscosity) change as the density 
profile evolves.  
The diffusivities scale as
$\chi_t, \nu_t \propto \sqrt{g L_z^3 (\Delta S_0 / c_P) / \text{Ra}_t}$.
Defining the thermal diffusion timescale as $t_\chi \equiv \tau_IL_z^2 / \chi$, the
ratio of $t_\chi$ to the buoyancy time is
\begin{equation}
\frac{t_\chi}{t_b} =\text{Ra}_t^{1/2}.
\label{eqn:timescales}
\end{equation}
We carry out two experiments in this study. In the first,
we fix $\epsilon$ and increase Ra, thus increasing the ratio in
(\ref{eqn:timescales}). In the second, we fix Ra and vary $\epsilon$,
scaling the dynamical timescales $(t_b, t_\chi)$ as $\epsilon^{-1/2}$
relative to the speed of sound; we see this reflected in the evolved Mach
number scaling (Fig. \ref{fig:ma_v_eps}).

We use $\ln \rho$ and $T$ as our thermodynamic variables and solve the Fully 
Compressible Navier-Stokes equations,
\begin{align}
&\begin{aligned}
&\frac{\partial \ln\rho}{\partial t} + \grad\cdot\bm{u} 
    = -\bm{u}\cdot\grad\ln\rho,
	\label{eqn:continuity_eqn}
\end{aligned}\\
&\begin{aligned}
\frac{\partial\bm{u}}{\partial t} + \grad T - 
&\nu\grad\cdot\lilstressT - \lilstressT\cdot\grad\nu =
-\bm{u}\cdot\grad\bm{u} - T\grad\ln\rho + \bm{g} + 
\nu\lilstressT\cdot\grad\ln\rho,
\label{eqn:momentum_eqn}
\end{aligned}\\
&\begin{aligned}
\frac{\partial T}{\partial t} -\frac{1}{c_V}\left(\right.\chi&\left.
    \grad^2 T + \grad T\cdot\grad\chi\right) =
	-\bm{u}\cdot\grad T - (\gamma-1)T\grad\cdot{\bm{u}}
	+ \frac{1}{c_V}\left(\chi\grad T \cdot\grad\ln\rho +
	\nu\left[\lilstressT\cdot\nabla\right]\cdot\bm{u}\right), 
	\label{eqn:energy_eqn}
\end{aligned}
\end{align}
with the viscous stress tensor given by
\begin{equation}
\sigma_{ij} \equiv \left(\frac{\partial u_i}{\partial x_j} + 
\frac{\partial u_j}{\partial x_i} - \frac{2}{3}\delta_{ij}\grad\cdot\bm{u}\right),
	\label{eqn:stress_tensor}
\end{equation}
where $\delta_{ij}$ is the Kronecker delta. Taking an inner product of
(\ref{eqn:momentum_eqn}) with $\rho\bm{u}$ and adding it to 
$\rho c_V\times$(\ref{eqn:energy_eqn}) reveals the full energy equation,
\begin{equation}
\frac{\partial}{\partial t}\left(\rho\left[\frac{|\bm{u}|^2}{2} + c_V T + \phi\right]\right) +
\Div{\bm{F}_{\text{conv}} + \bm{F}_{\text{cond}}} = 0,
	\label{eqn:energy_eqn_full}
\end{equation}
where
$
\bm{F}_{\text{conv}} \equiv \bm{F}_{\text{enth}} + \bm{F}_{\text{KE}} + \bm{F}_{\text{PE}} + \bm{F}_{\text{visc}}
$
is the convective flux and $\bm{F}_{\text{cond}} = -\kappa \grad T$
is the conductive flux.
The individual contributions to $\bm{F}_{\text{conv}}$ are the enthalpy flux, 
$\bm{F}_{\text{enth}} \equiv \rho\bm{u}(c_V T + P/\rho)$;
the kinetic energy flux, 
$\bm{F}_{\text{KE}} \equiv \rho|\bm{u}|^2\bm{u}/2$;
the potential energy flux,
$\bm{F}_{\text{PE}} \equiv \rho\bm{u}\phi$ (with $\phi \equiv -gz$);
and the viscous flux, 
$\bm{F}_{\text{visc}} \equiv -\rho\nu\bm{u}\cdot\lilstressT$.
Understanding how each of these fluxes interact  
is crucial in characterizing convective heat transport.

We utilize the 
Dedalus\footnote{\url{http://dedalus-project.org/}} 
pseudospectral framework \cite{burns&all2016} to time-evolve  
(\ref{eqn:continuity_eqn})-(\ref{eqn:energy_eqn}) 
using an implicit-explicit (IMEX), third-order, four-step 
Runge-Kutta timestepping scheme RK443 \cite{ascher&all1997}.  
Thermodynamic variables are decomposed such that $T = T_0 + T_1$ and
$\ln\rho = (\ln\rho)_0 + (\ln\rho)_1$, 
and the velocity is $\bm{u} = w\bm{\hat{z}} + u\bm{\hat{x}} + v\bm{\hat{y}}$.
In our 2D runs, $v = 0$.
Subscript 0 variables, set by (\ref{eqn:polytrope}), 
have no time derivative and vary only in $z$.
Variables are time-evolved on a dealiased Chebyshev (vertical)
and Fourier (horizontal, periodic) domain in which the
physical grid dimensions are 3/2 the size of the coefficient grid.  
Domain sizes range from
64x256 coefficients at the lowest values of 
Ra to 1024x4096 coefficients at Ra $> 10^{7}$ in 2D,
and from 64x128$^2$ to  256x512$^2$ in 3D. 
By using IMEX timestepping, we implicitly step the 
stiff linear acoustic wave contribution and are able to
efficiently study flows at high ($\sim 1$) 
and low ($\sim 10^{-4}$) Ma.  Our equations take the form
of the FC equations in \cite{lecoanet&all2014}, extended to include
$\nu$ and $\chi$ which vary with depth, and we follow the approach there.
This IMEX approach has been successfully 
tested against a nonlinear benchmark  of the compressible 
Kelvin-Helmholtz instability \cite{Lecoanet_et_al_2016_KH}.

We impose impenetrable, stress free, fixed temperature boundary conditions at
the top and bottom of the domain, with 
$w = \partial_z u = T_1 = 0$ at $z = \{0, L_z\}$. 
$T_1$ is initially filled with
random white noise whose magnitude is infinitesimal
compared to $\epsilon T_0$.
We filter this noise spectrum in coefficient space, 
such that only the lower 25\% of the coefficients
have power. All reported results are taken from time averages
over many $t_b$ beginning \{100, 40\}$t_b$
after the start of our \{2D, 3D\} simulations to
ensure our results are not biased by the convective transient.

\section{Results \& Discussion}
\label{sec:results}

Solutions were time-evolved until a long time average of the fluxes
showed little
variance with depth. A linear stability analysis determined
that convective onset
occurs at $\text{Ra}_{\text{crit}} = \{11.15, 10.06, 10.97, 10.97\}$ 
for $\epsilon = \{1.0, 0.5, 10^{-4}, 10^{-7}\}$, respectively.

\vspace{-0.5cm}
\subsection{Evolved fluid numbers \& flow morphology}

\begin{figure}[t!]
\includegraphics[width=\textwidth]{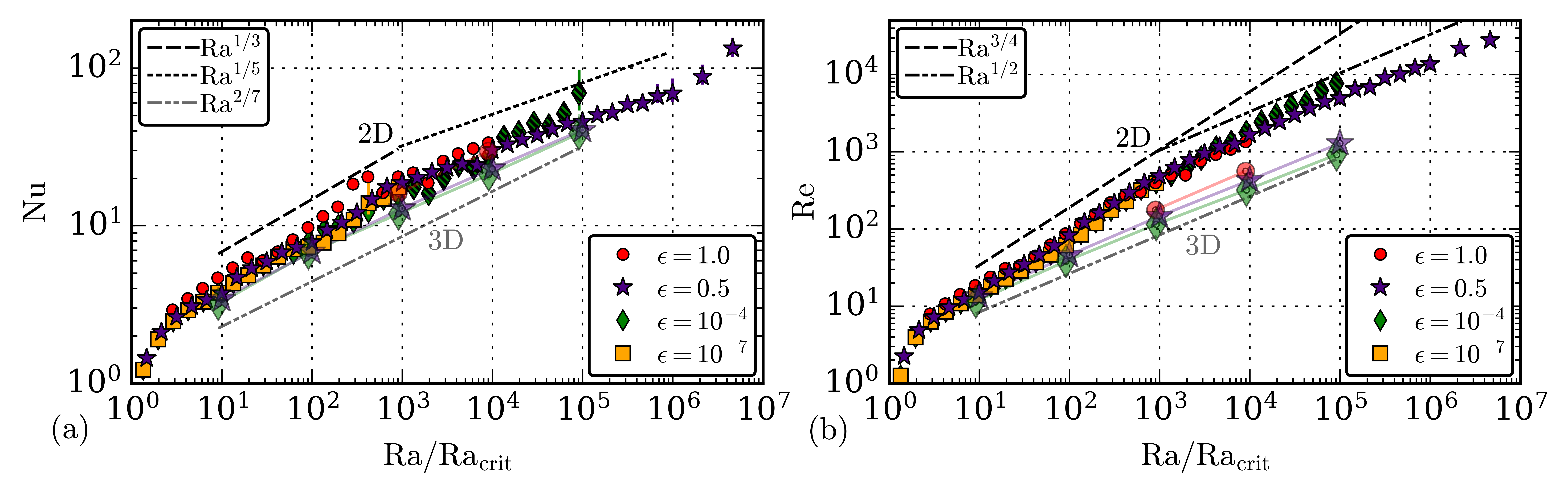}
\caption{
Flow properties at high and low $\epsilon$. 
(a) Nu vs. Ra/Ra$_{\text{crit}}$.
Errors bars indicate the variance of Nu with depth;
large error bars indicate a poorly converged solution.
(b) Re vs. Ra/Ra$_{\text{crit}}$.
Re is measured at the midplane of the atmosphere.
Larger symbols with inlaid circles designate 3D runs.
 \label{fig:re_and_nu_v_ra}
}
\end{figure}

We measure the adiabatic Mach number (Ma = $|\bm{u}|/\sqrt{\gamma T}$),
and find that it is a strong function of 
$\epsilon$ and a weak function of Ra.  
In our 2D runs, when Ma $< 1$, we observe a scaling law of
Ma(Ra$, \epsilon) \appropto \epsilon^{1/2}$Ra$^{1/4}$.
This relation breaks down as the mean
Ma approaches 1 (Fig. \ref{fig:ma_v_eps}).  This transition
occurs near Ra/Ra$_\text{crit} \approx \{10^{2}, 10^{3}\}$ for $\epsilon = \{1, 0.5\}$.
In our limited 3D runs, Ma appears to be a function of $\epsilon$ alone, with
Ma $\appropto \epsilon^{1/2}$, so at high Ra, Ma$_{\text{3D}} < \text{Ma}_{\text{2D}}$.
We conjecture that the scaling of Ma with Ra in the 2D runs 
is due to the formation of coherent high-velocity ``spinners,'' which form
between upflow and downflow lanes.  These structures, which are reminiscent of flywheel
modes in RBC, do not appear in our 3D runs at these parameters
\cite{jones&all1976, brummell&all2002}.
Simulations in the range of Ra/Ra$_{\text{crit}} > 10^3$ at $\epsilon = 10^{-4}$
exhibited ``windy'' states of convection, in which a large-scale horizontal
shearing flow replaced the more standard upflow/downflow morphology of
convection.  Similar states have been studied in
RBC \cite{goluskin&all2014}.  These runs are represented in Figs. 
\ref{fig:ma_v_eps}, \ref{fig:re_and_nu_v_ra}, \& \ref{fig:nrho_v_ra}
as hatched points, and while this phenomenon does not appear to greatly modify the
scaling of fluid properties measured in this work, these states warrant
further investigation.

\begin{figure}[t!]
\includegraphics[width=\textwidth]{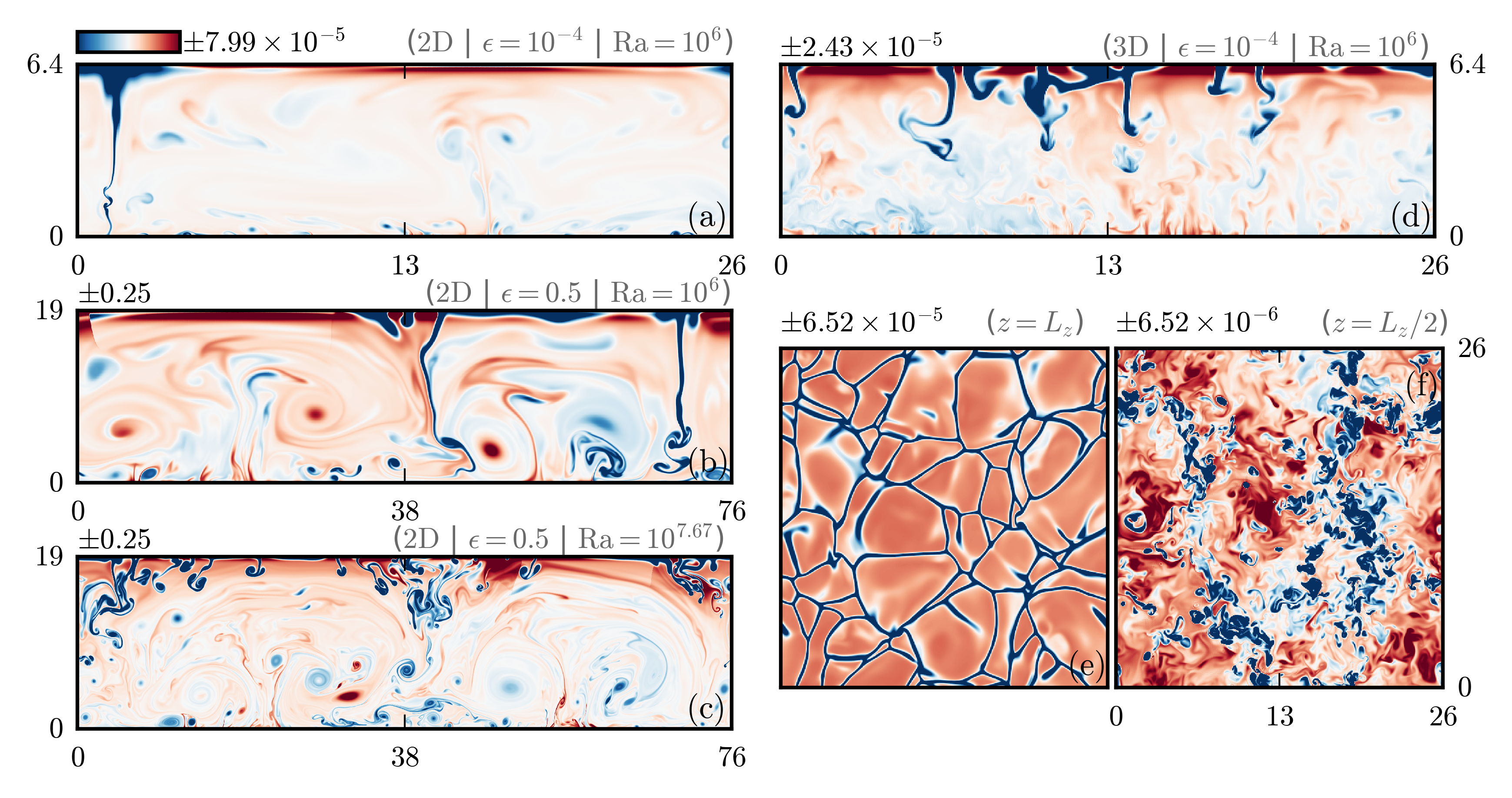}
\caption{Characteristic entropy fluctuations in evolved flows roughly
$\{$140, 60$\}$ $t_b$ after the start of simulations for $\{$2D, 3D$\}$ runs. 
The time- and horizontally-averaged profile is removed in vertical slices (a)-(d). 
The time- averaged mean value is removed in horizontal slices (e) and (f).
(a) A low Ma flow at moderate Ra. (b) A high Ma flow at the same Ra as in (a).
(c) A high Ma flow at high Ra. 
Shock systems can be seen in the upper atmosphere of the high Ma flows,
for example at $(x, z) \sim (5, 15-19)$ in (b) and $(x, z) \sim 
(50, 15-19)$ in (c).
(d)-(f) A low Ma 3D run at the same parameters as in (a),
where (d) is a vertical ($x$, $z$) slice at $y = L_y/2$, (e) is a horizontal slice
at $z = L_z$, and (f) is a horizontal slice at $z = L_z/2$.
\label{fig:entropy_snapshots} }
\end{figure}

In 2D, low Ma flows (e.g., $\epsilon = 10^{-4}$)
display the classic narrow downflow and broad upflow lanes of stratified
convection (Fig. \ref{fig:entropy_snapshots}a).
At high Ma (e.g., $\epsilon = 0.5$, Ra/Ra$_{\text{crit}} \gtrsim 10^3$), 
bulk thermodynamic structures are similar but
shock systems form in the upper atmosphere near downflow lanes 
(Fig. \ref{fig:entropy_snapshots}b\&c), as reported previously
\cite{cattaneo&all1990, malagoli&all1990}.
At large Ra, the diffusion timescale becomes long (\ref{eqn:timescales}), 
and
thermodynamic structures form small eddies which traverse the
domain repeatedly before diffusing (Fig. \ref{fig:entropy_snapshots}c).
As evidenced by the colorbar scalings, the
amplitudes of thermodynamic fluctuations scale with $\epsilon$.

In 3D, the same upflow/downflow asymmetry is seen, but other aspects 
of the flow are distinctly different.  Fig. \ref{fig:entropy_snapshots}d-f show select
snapshots of a 3D simulation with the same input parameters as the 2D case in Fig. \ref{fig:entropy_snapshots}a.
In 2D, large-scale, coherent 
spinners dominate the  flow, leading to a single upflow and downflow.  
New downflowing plumes at the upper boundary are efficiently swept into the large 
coherent structure (near $x \sim 1$ and spanning the vertical domain).  The 
behavior of downflows in 3D is strikingly different (Fig. \ref{fig:entropy_snapshots}d).
In 3D, many individual plumes detach from 
the upper boundary, but do not organize into a single dominant downflow 
in the same fashion.  Horizontal cuts near the top of the domain 
(Fig. \ref{fig:entropy_snapshots}e) reveal 
a network of narrow downflow lanes surrounding broad upflows. 
Stronger clusters of downflows near the surface are linked to 
sheets of low entropy at the midplane of the domain (Fig. \ref{fig:entropy_snapshots}f).
As the flows evolve in time, new downflows appear at the top of the domain in 
the middle of upflows and join the surrounding downflow network, 
causing the convective structures to fragment.  There is no preferred orientation 
in the newly forming downflows, and the convective flow field constantly evolves, 
which appears to prevent the occurrence of either spinners or windy states.

The efficiency of convection is quantified by the Nusselt number (Nu).  
Nu is well-defined in RBC
as the total flux normalized by the steady-state conductive flux 
\cite{johnston&doering2009, otero&all2002}.
In stratified convection Nu is more difficult to define, and we use
a modified version of a traditional stratified Nusselt number 
\cite{graham1975,hurlburt&all1984},
\begin{equation}
\text{Nu} \equiv \frac{\langle F_{\text{conv,z}} + F_{\text{cond,z}} - F_{\text{A}}\rangle}
{\langle F_{\text{cond,z}} - F_{\text{A}}\rangle} 
= 1 + \frac{\langle F_{\text{conv,z}}\rangle}{\langle F_{\text{cond,z}} - F_{\text{A}} \rangle}
\label{eqn:nusselt}
\end{equation}
where $F_{\text{conv,z}}$ and $F_{\text{cond,z}}$ are the 
z-components of $\bm{F}_{\text{conv}}$ and $\bm{F}_{\text{cond}}$,
and $\langle \rangle$ are volume averages.  
$F_{\text{A}} \equiv -\langle\kappa\rangle \partial_z T_{\text{ad}}$ 
is the conductive flux of the proper corresponding adiabatic atmosphere.
For a compressible, ideal gas in hydrostatic equilibrium,
$\partial_z T_{\text{ad}} \equiv - g / c_{P}$ \cite{spiegel&veronis1960}.  
It is important to measure the evolved value of
$\langle \kappa \rangle = \langle \rho\chi \rangle$, which is nearly
$\kappa_0$ for small $\epsilon$ but changes appreciably for large
values of $\epsilon$.
In incompressible Boussinesq convection, where $\grad S = 0$ only when 
$\grad T = 0$, and where $\kappa$ is constant with depth and time,
this definition reduces to the traditionally defined
Nusselt number \cite{otero&all2002, johnston&doering2009}.

The variation of Nu with Ra is shown in 
Fig. \ref{fig:re_and_nu_v_ra}a.  We find that the Nu depends primarily
on Ra, not on $\epsilon$, except where dynamical regimes change.
In 2D and at low to moderate Ra, 
Nu $\appropto$ Ra$^{1/3}$ regardless of $\epsilon$,
reminiscent of scaling laws in \RB boundary layer theory 
\cite{grossman&lohse2000, ahlers&all2009, king&all2012}.
As the flow becomes supersonic,  Nu $\appropto$ Ra$^{1/5}$.
It is also important to note that, in 2D,
the value of Nu is heavily dependent upon the specific thermodynamic
structures of the solution, and slight changes in
Ra can result in a simulation converging to one solution or another. 
Select simulations were run at higher aspect ratios (8 and 16), and similar flow
morphologies were obtained, suggesting that these states are not highly
sensitive to aspect ratio.
In our limited 3D runs, it appears that Nu $\appropto$ Ra$^{2/7}$, a classic scaling law
seen in RBC \cite{johnston&doering2009}.

The rms Reynolds number (Re = $|\bm{u}|L_z/\nu$) and Peclet number
(Pe = Pr Re)
compare the importance of advection to diffusion in the evolved
convective state.  For Pr = 1, Pe = Re.  
Our choice of $\{\nu,\chi\}\propto \rho_0^{-1}$ drastically changes
the value of Re between the top and bottom of the atmosphere.  We report values of
Re at the midplane ($z=L_z/2$) of the atmosphere in
Fig. \ref{fig:re_and_nu_v_ra}b.  We find that Re
depends largely on Ra, but not $\epsilon$, except when the flow regime
changes.
In 2D Re $\appropto$ Ra$^{3/4}$ at low Ra.    When the 2D flows
become supersonic, 
Re $\appropto$ Ra$^{1/2}$, as expected
from (\ref{eqn:timescales}).
In our limited 3D runs,
Re $\appropto$ Ra$^{1/2}$, consistent with the supersonic results.
The heightened scaling
of Re in 2D follows the scaling of velocity (Ma) with Ra, as
seen in Fig. \ref{fig:ma_v_eps}, and reflects the presence of coherent
spinners, which do not exist in 3D.

\vspace{-0.5cm}
\subsection{Evolved stratification}

\begin{figure}[t!]
\includegraphics[width=\textwidth]{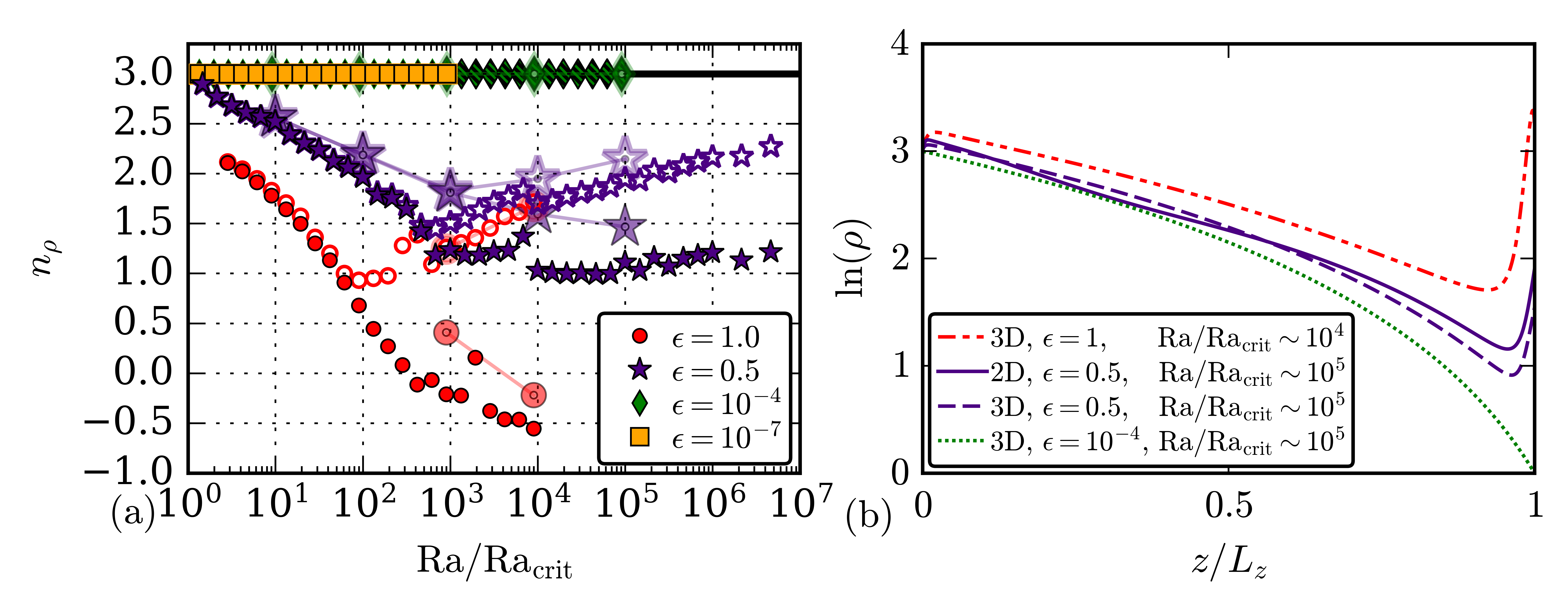}
\caption{\label{fig:nrho_v_ra} 
(a) Solid symbols show the density contrast measured
in density scale heights between the upper and lower boundary, 
$n_\rho = \ln[\rho(z=0)/\rho(z=L_z)]$.
Empty symbols show 
$n_\rho = \ln[\text{max}(\rho)/\text{min}(\rho)]$. 
At low $\epsilon$ the evolved
$n_{\rho}$ is close to the initial conditions of $n_\rho = 3$.  
At high $\epsilon$,
the density stratification decreases.  Once the mean 
Ma approaches 1 (at Ra/Ra$_{\text{crit}} \approx \{10^2, 10^3\}$ for $\epsilon = \{1, 0.5\}$
as in Fig. \ref{fig:ma_v_eps}b), density inversions form within the thermal
boundary layers. Larger symbols represent 3D runs. (b) The natural log of time-
and horizontally-averaged density profiles are shown for select simulations,
to illustrate the inversions which appear in the boundary layers.}
\end{figure}

In the evolved state, the flows can change the density stratification.
In Fig. \ref{fig:nrho_v_ra}a, we measure the 
time- and horizontally-averaged density profile in two ways. Empty symbols
show the number of density scale heights between the maximum and minimum of the
atmospheric density profile.  Solid symbols
show the number of density scale heights between the top and bottom of the atmosphere. 
We find that near-sonic and supersonic flows support significant, 
persistent density inversions in the boundary layers, as
was reported previously \cite{brandenburg&all2005}.  This is visible when
solid symbols lie below empty symbols.  We find this in 2D and 3D, even
at very large $\epsilon$.  

Sample evolved density profiles are displayed in Fig. \ref{fig:nrho_v_ra}b.
The natural log of the temporally and horizontally averaged density profile, 
$\ln \rho = \ln \rho_0 + \ln\rho_1$, is displayed for four cases.  At low $\epsilon$
(dotted green line), the density stratification is, to first order, unchanged from
the initial density stratification.  At high $\epsilon$, in both 2D (solid purple line)
and 3D (dashed purple line and dash-dot-dot red line), the evolved stratification differs
significantly from the initial state and does not increase monotonically with depth.
To measure the number of density scale heights between two points in the atmosphere,
$z_1$ \& $z_2$, we calculate $n_\rho(z_1, z_2) = \ln\rho(z_2) - \ln\rho(z_1)$.  Thus,
the values plotted in Fig. \ref{fig:nrho_v_ra}a for the cases in Fig. \ref{fig:nrho_v_ra}b
can be directly read off.  For example, at $\epsilon = 1$ and Ra/Ra$_{\text{crit}} \sim 10^4$
(dash-dot-dot red line), measuring the stratification between the boundaries retrieves
$n_\rho(L_z, 0) \approx -0.3$, but measuring between the maximum and minimum value of
the profile retrieves $n_\rho(\text{min, max}) \approx 1.6$.

Surprisingly, the evolved $n_\rho$ is always less than the initial $n_\rho = 3$,
and turbulent pressure support plays a larger role than atmospheric slumping.
This appears to arise as a result of convection making the interior isentropic
in the presence of fixed-temperature boundary conditions;
we expect the behavior of the stratification to be dependent on the choice of
thermal boundary conditions.
The agreement of Nu \& Re across $\epsilon$ (Fig. \ref{fig:re_and_nu_v_ra}), 
particularly at low Ra in which all four of our cases collapse onto a single
power law, is striking in light of the vastly different evolved stratifications
felt by the flows.

\vspace{-0.5cm}
\section{Discussion \& Future Work}
We have found that the evolved flow properties of stratified,
compressible convection scale in a manner reminiscent of incompressible,
Boussinesq \RB convection.
We argue that polytropically stratified atmospheres are the natural
extension of the RBC problem with an additional control parameter, $\epsilon$,
whose primary role is to set the Ma of the flows.  We show that other properties
of the evolved solutions (Nu, Re) are nearly identical at vastly different values
of $\epsilon$, except for where there is a transition between the subsonic
and supersonic regimes.  We also see that Nu scales similarly in 2D and 3D,
and that Ma in 3D solutions appears to be a function of $\epsilon$ alone,
allowing for simple specification of the evolved Ma using input parameters.
The stratification of 
these polytropic atmospheres evolves in a complex
manner, and future work should aim to 
understand the importance of stratification on
convective heat transport and other flow properties.

Our studies here will serve as a foundation for 
comparing heat transport in stratified convection
to that in RBC \cite{johnston&doering2009}
and for better quantifying transport in stratified convection.  
These results can be used to determine if fluid properties
scale appropriately in simplified equation sets, 
such as the anelastic equations.
This work will also be useful in coming to understand more realistic systems, 
such as rapidly rotating atmospheres \cite{julien&all2012},
atmospheres bounded by stable regions \cite{hurlburt&all1986}, 
and regions with realistic profiles of $\kappa$.

\begin{acknowledgments}
EHA acknowledges the support of the University of Colorado's George 
Ellery Hale Graduate Student Fellowship.
This work was additionally supported by  NASA LWS grant number NNX16AC92G.  
Computations were conducted 
with support by the NASA High End Computing (HEC) Program through the NASA 
Advanced Supercomputing (NAS) Division at Ames Research Center on Pleiades
with allocations GID s1647 and GID g26133.
We thank Jon Aurnou, Axel Brandenburg, Keith Julien, Mark Rast, and Jeff Oishi 
for many useful discussions. We also thank the three anonymous referees whose
careful comments greatly improved the quality of this paper.
\end{acknowledgments}

\appendix
\section{Table of simulation parameters}
In table \ref{table:run_info}, we report both the input parameters ($\epsilon$, Ra$_t$, resolution)
and output values (evolved Nu, Re, Ma, $n_\rho$) of select simulations.  All 3D simulations
are listed, and the corresponding 2D simulations at the same parameters are included.  A full
table of all information is included as a CSV file in the supplemental materials.

\begin{center}
\begin{longtable}{ p{1cm} p{1cm} p{1cm} p{1cm} p{1cm} p{1cm} p{1.75cm} p{1.5cm} p{2.75cm} p{1cm} p{1.2cm}  }
\caption{Input simulation parameters of $\epsilon$, Ra, and resolution are given for
select simulations.
Output values of Nu, Ma, Re, and $n_\rho$, as plotted throughout the paper, are 
provided for the corresponding run.  The mean of Nu and Ma is reported, as well
as the distance from the mean to the atmospheric maximum and minimum.  Re is reported
at the midplane.  The evolved $n_\rho$ is reported using the two metrics described
in Fig. \ref{fig:nrho_v_ra}. \label{table:run_info}}\\
\hline
$\epsilon$	&	Ra$_t$	&	3D	&	$n_z$	&	$n_x$	&	$n_y$	&	Nu					&	Re	&	Ma					&	$n_{\rho, \text{max}}$	&	$n_{\rho, \text{bounds}}$	\\
\hline\hline \endfirsthead																													\\
\multicolumn{11}{c}%
{ \tablename\ \thetable{} -- continued from previous page} \\
\hline
$\epsilon$	&	Ra$_t$	&	3D	&	$n_z$	&	$n_x$	&	$n_y$	&	Nu					&	Re	&	Ma					&	$n_{\rho, \text{max}}$	&	$n_{\rho, \text{bounds}}$	\\
\hline \hline \endhead
\hline \endfoot
\hline \endlastfoot

\vspace{0.08cm}1	&	$10^4$	&	No	&	256	&	1024	&	---	&$	20.43	_{-	0.21	}^{+	0.24	}$&	397.69	&$	0.79	_{-	0.35	}^{+	5.49	}$&	1.26	&	-0.21	\\
\vspace{0.08cm}1	&	$10^5$	&	No	&	512	&	2048	&	---	&$	33.60	_{-	0.41	}^{+	1.03	}$&	1356.44	&$	0.82	_{-	0.33	}^{+	4.51	}$&	1.72	&	-0.55	\\
\vspace{0.08cm}0.5	&	100	&	No	&	64	&	256	&	---	&$	3.72	_{-	0.01	}^{+	0.01	}$&	15.22	&$	0.22	_{-	0.11	}^{+	0.33	}$&	2.52	&	2.52	\\
\vspace{0.08cm}0.5	&	1000	&	No	&	128	&	512	&	---	&$	7.87	_{-	0.04	}^{+	0.09	}$&	83.02	&$	0.39	_{-	0.26	}^{+	0.71	}$&	1.98	&	1.96	\\
\vspace{0.08cm}0.5	&	$10^4$	&	No	&	128	&	512	&	---	&$	18.69	_{-	0.38	}^{+	0.63	}$&	487.35	&$	0.74	_{-	0.54	}^{+	1.50	}$&	1.51	&	1.24	\\
\vspace{0.08cm}0.5	&	$10^5$	&	No	&	256	&	1024	&	---	&$	30.14	_{-	0.15	}^{+	0.39	}$&	1677.07	&$	0.71	_{-	0.34	}^{+	0.87	}$&	1.70	&	1.03	\\
\vspace{0.08cm}0.5	&	$10^6$	&	No	&	512	&	2048	&	---	&$	45.86	_{-	1.90	}^{+	3.76	}$&	4943.73	&$	0.64	_{-	0.24	}^{+	0.62	}$&	1.95	&	1.11	\\
\vspace{0.08cm}$10^{-4}$	&	100	&	No	&	64	&	256	&	---	&$	3.75	_{-	0.01	}^{+	0.01	}$&	13.93	&$	2.82	_{-	1.39	}^{+	1.91}\cdot 10^{-3}	$&	3.00	&	3.00	\\
\vspace{0.08cm}$10^{-4}$	&	1000	&	No	&	128	&	512	&	---	&$	7.97	_{-	0.07	}^{+	0.28	}$&	65.85	&$	4.51	_{-	2.66	}^{+	3.95}\cdot 10^{-3}	$&	3.00	&	3.00	\\
\vspace{0.08cm}$10^{-4}$	&	$10^4$	&	No	&	128	&	512	&	---	&$	15.21	_{-	0.14	}^{+	0.32	}$&	391.38	&$	9.30	_{-	7.11	}^{+	11.7}\cdot 10^{-3}	$&	3.00	&	3.00	\\
\vspace{0.08cm}$10^{-4}$	&	$10^5$	&	No	&	256	&	1024	&	---	&$	27.83	_{-	0.47	}^{+	0.42	}$&	1766.70	&$	1.50	_{-	1.17	}^{+	2.19}\cdot 10^{-2}	$&	3.00	&	3.00	\\
\vspace{0.08cm}$10^{-4}$	&	$10^6$	&	No	&	512	&	2048	&	---	&$	69.68	_{-	15.80	}^{+	28.24	}$&	7684.29	&$	2.10	_{-	1.74	}^{+	3.15}\cdot 10^{-2}	$&	3.00	&	3.00	\\
\vspace{0.08cm}1	&	$10^4$	&	Yes	&	128	&	256	&	256	&$	16.10	_{-	0.99	}^{+	0.15	}$&	176.84	&$	0.39	_{-	0.13	}^{+	3.78	}$&	1.24	&	0.41	\\
\vspace{0.08cm}1	&	$10^5$	&	Yes	&	256	&	512	&	512	&$	29.67	_{-	1.90	}^{+	0.44	}$&	562.40	&$	0.38	_{-	0.14	}^{+	3.71	}$&	1.67	&	-0.22	\\
\vspace{0.08cm}0.5	&	100	&	Yes	&	64	&	128	&	128	&$	3.42	_{-	0.11	}^{+	0.06	}$&	13.42	&$	0.21	_{-	0.08	}^{+	0.33	}$&	2.56	&	2.57	\\
\vspace{0.08cm}0.5	&	1000	&	Yes	&	128	&	256	&	256	&$	6.83	_{-	0.25	}^{+	0.18	}$&	46.03	&$	0.23	_{-	0.08	}^{+	0.38	}$&	2.19	&	2.19	\\
\vspace{0.08cm}0.5	&	$10^4$	&	Yes	&	128	&	256	&	256	&$	12.89	_{-	0.62	}^{+	0.30	}$&	146.92	&$	0.23	_{-	0.08	}^{+	0.45	}$&	1.83	&	1.81	\\
\vspace{0.08cm}0.5	&	$10^5$	&	Yes	&	256	&	512	&	512	&$	23.05	_{-	1.04	}^{+	1.41	}$&	429.63	&$	0.21	_{-	0.08	}^{+	0.43	}$&	1.95	&	1.59	\\
\vspace{0.08cm}0.5	&	$10^6$	&	Yes	&	256	&	512	&	512	&$	40.53	_{-	0.72	}^{+	0.66	}$&	1291.61	&$	0.20	_{-	0.08	}^{+	0.46	}$&	2.15	&	1.47	\\
\vspace{0.08cm}$10^{-4}$	&	100	&	Yes	&	64	&	128	&	128	&$	3.19	_{-	0.10	}^{+	0.06	}$&	11.48	&$	2.53	_{-	9.97	}^{+	1.93}\cdot 10^{-3}	$&	3.00	&	3.00	\\
\vspace{0.08cm}$10^{-4}$	&	1000	&	Yes	&	128	&	256	&	256	&$	6.72	_{-	0.19	}^{+	0.57	}$&	38.32	&$	2.88	_{-	1.09	}^{+	2.55}\cdot 10^{-3}	$&	3.00	&	3.00	\\
\vspace{0.08cm}$10^{-4}$	&	$10^4$	&	Yes	&	128	&	256	&	256	&$	11.92	_{-	0.40	}^{+	1.34	}$&	115.06	&$	2.85	_{-	1.13	}^{+	3.11}\cdot 10^{-3}	$&	3.00	&	3.00	\\
\vspace{0.08cm}$10^{-4}$	&	$10^5$	&	Yes	&	256	&	512	&	512	&$	21.11	_{-	1.69	}^{+	4.74	}$&	319.84	&$	2.46	_{-	1.07	}^{+	2.90}\cdot 10^{-3}	$&	3.00	&	3.00	\\
\vspace{0.08cm}$10^{-4}$	&	$10^6$	&	Yes	&	256	&	512	&	512	&$	38.50	_{-	4.71	}^{+	12.54	}$&	920.21	&$	2.26	_{-	1.02	}^{+	2.84}\cdot 10^{-3}	$&	3.00	&	3.00	\\
\hline
\end{longtable}
\end{center}

\end{document}